\documentclass[runningheads]{llncs}
\usepackage{graphicx}
\begin{document}
\title{Characterizing the Experience of Subjects in Software Engineering Studies}
%
%
\author{Rafael de Mello\inst{1}\orcidID{0000-0002-9877-3946} \and
Matheus Coelho\inst{1}} 
%
\authorrunning{de Mello and Coelho}
%
\institute{CEFET/RJ, Rio de Janeiro, Brazil \\
\email{rafael.mello@cefet-rj.br, matheus.coelho@aluno.cefet-rj.br}}

\maketitle              
\begin{abstract}
Context: Empirical studies in software engineering are typically centered on human subjects, ranging from novice to experienced developers. The experience of these individuals is a key context factor that should be properly characterized for supporting the design of empirical studies and interpreting their results. However, the criteria adopted for characterizing the experience of subjects do not follow a standard and are frequently limited. Goal: Our research aims at establishing an optimized and comprehensive scheme to characterize the subjects' experience for studies in software engineering. Method: Based on previous work, we defined the first version of this scheme, composed of three experience attributes, including time, number of projects, and self-perception. In the last years, we applied the characterization scheme over four empirical studies, reaching the characterization of 79 subjects in three different skills. Results: We found that the attributes from our scheme are positively but moderately correlated. This finding suggests these attributes play a complementary role in characterizing the subjects' experience. Besides, we found that study subjects tend to enumerate the technical diversity of their background when summarizing their professional experience. Conclusion: The scheme proposed represents a feasible alternative for characterizing subjects of empirical studies in the field. However, we intend to conduct additional investigations with developers to evolve it. 

\end{abstract}
%
%
\section{Introduction}
Software Engineering is a sociotechnical field composed of several semi-automated and manual activities strongly influenced by human aspects at different levels~\cite{lenberg2015behavioral}. Consequently, empirical studies in software engineering typically require the involvement of humans as subjects, ranging from novice students to experienced practitioners~\cite{de2015characterizing}\cite{falessi2018empirical}. Empirical studies centered on practitioners include not only controlled studies but also studies intensive on qualitative data, such as opinion surveys~\cite{linaker2015guidelines}, focus group sessions~\cite{kontio2008focus}, case studies ~\cite{runeson2009guidelines}, and action research~\cite{dos2009action}. In most of these studies, it is common to gather some characterization data about the study subjects. In this way, characterizing their experience is often recognized as a relevant but also challenging task ~\cite{host2005experimental}\cite{de2016surveys}\cite{falessi2018empirical}. 

Among other utilities, characterizing subjects' experience is a key resource to support the design of empirical studies and to support the interpretation of their findings~\cite{host2005experimental}~\cite{de2015characterizing}. For instance, researchers may use experience data to compose balanced pairs of developers for assigning code review tasks in the context of a controlled study~\cite{oliveira2017collaborative}. This experience data may be also used to assess to which extent the experience in code reviews influenced the performance of each pair. 




Despite the importance of properly characterizing the subjects' experience, one can see that characterization data gathered in empirical studies is frequently insufficient or underused, which may be partially explained by the lack of standardization of the characterization instruments used in the field. For instance,~\cite{de2015characterizing} identified that surveys in software engineering typically characterize subjects by few and frequently disconnected attributes from the perspective of the research topic. Consequently, characterization data- including experience- is barely used in the interpretation of the results. 
Besides, we argue that an inadequate characterization of the subjects' experience may lead to misinterpretations of the studies' results. 


In this sense, one risky strategy is relying on a single attribute for characterizing experience. For instance, let us consider two co-workers named Johnny and Selena. Johnny has ten years of experience in reviewing code, while Selena has six years. Selena had worked reviewing code in a diverse portfolio of 20 projects, while Johnny worked with code review in 10 projects. When individually asked about their background, Johnny understands that he has more experience with code reviews than Selena. Besides, Selena understands that she has less experience in code reviews than Johnny. Now, let us consider that both professionals are volunteers in a controlled study on code reviews involving several other developers. If the researchers opted by characterizing the subjects' experience in code reviews only based on the number of projects, Johnny would be ranked under Selena. However, if the experience is grounded in self-assessment or years, Johnny would be ranked above Selena. 


Although we should recognize that characterizing and comparing subjective context factors is challenging, the aforementioned example illustrates how a more comprehensive characterization, considering different perspectives, would improve interpreting the subjects' experience. However, one may see that researchers should avoid long characterization forms that would discourage the participation of subjects. In this paper, we propose the first version of a lightweight scheme for characterizing the experience of subjects from empirical studies in software engineering. This scheme is composed of three easy-to-assess attributes of experience: self-assessment (Likert scale), time (years), and number of projects. We also report in this paper a large-scale investigation on the feasibility of the proposed scheme. In the last five years, we applied a standard questionnaire based on this scheme to characterize the experience of 79 subjects from four empirical studies. In these studies, we used the scheme attributes to characterize the subjects' experience in software development and two more specific activities. Besides, the studies' subjects were asked to summarize their professional experience in software development.

The study results reveal that the scheme attributes are positively but moderately correlated. This finding suggests that researchers may combine these attributes for reaching a more comprehensive and reliable characterization of the developers' experience. Besides, we also found that developers tend to self-describe their experience based on other attributes addressing the diversity of their technical experience, such as enumerating the development activities performed and development technologies employed. Based on these findings, we conclude that the proposed scheme is feasible although we could improve its comprehensiveness. In this way, we are planning an in-depth investigation of the beliefs and values of researchers and developers about experience.

Section~\ref{sec:relatedwork} presents the related work, in which we discuss approaches for characterizing the subjects' experience. Section~\ref{sec:design} describes the settings of our feasibility study, also introducing the first version of our proposed scheme. Section~\ref{sec:results} presents the results of our study, reporting its main findings. Section~\ref{sec:discussion} presents a discussion of the study findings, reflecting on the strengths of the scheme proposed and possible opportunities for evolving it. Section~\ref{sec:threats} discusses threats to validity. Finally, Section~\ref{sec:conclusion} concludes the paper and indicates future work.


\section{Related Work}
\label{sec:relatedwork}

As far as we are aware, our research is the first one assessing and comparing different perspectives for characterizing the experience of subjects in studies from the field. However, there are several related work addressing different aspects involved in the research motivation and design choices. 

In \cite{dybaa2012works}, the authors discusses the role of context on interpreting the results of empirical studies from the field. In this way, they mapped a scheme in composed of omnibus dimensions and discrete context. Among the omnibus dimensions, the authors highlight the importance of properly characterizing the study participants. For this purpose, they exemplify some context factors used in empirical studies. However, none of them address the characterization of subjects' experience. In this way, one can see that merely labeling participants as students or professionals is insufficient. In~\cite{host2000using}, the authors investigated to which extent the performance of subjects ranked as students and professional developers are different. Based on their findings, the authors concluded that software engineering students may be used in empirical studies instead of professional software developers under certain conditions. In a more recent study~\cite{salman2015students}, the authors recruited a group of students and a group of professionals for performing test-driven development tasks. Then, the authors measured and compared the quality of the tasks performed by these groups. As a result, they concluded that was no group reached better results than the other. 

We also found related work proposing schemes for characterizing subjects in empirical studies. In \cite{host2005experimental}, the authors proposed a scheme for characterizing subjects through their motivation (incentive) and experience. For measuring experience, the scheme rank subjects in five levels by combining their academic level with their time working in industrial projects. However, the working time is grounded in only three arbitrary ranges: less than three months, between three months and two years, and more than two years. After applying the scheme over the characterization data of previous experiments, the authors found a good level of agreement among the original classification and those resulting from the proposed scheme. Here, it is important to note that reaching high levels of agreement does not assure that both classifications are right. In other words, eventual misclassifications in the original studies may have been replicated or even intensified. Besides, one can see that the proposed scheme is insufficient to distinguish more experienced groups of subjects once any individual having more than two years of professional experience is considered highly experienced.

More recently, Falessi et al.~\cite{falessi2018empirical} discussed the importance of empirical studies in software engineering going beyond a basic distinction between students and professionals, arguing that characterizing the subjects' experience is a key strategy for interpreting the results of empirical studies. After conducting qualitative studies with Empirical Software Engineering specialists, the authors proposed the $R^3$ scheme for characterizing the experience of subjects. This scheme advocates that experience should be characterized from from three perspectives: the subjects' \textit{real} experience, the subjects' \textit{relevant} experience and the \textit{recent} experience. Although no formal set of experience attributes are defined, the authors exemplify the measurement of the different perspectives of experience by measuring the number of years of experience. However, they suggest using unbalanced intervals of years for ranking the level of experience. Besides, the authors also suggest arguing subjects about their participation in projects (industry, academic, open-source) as an additional criterion for assessing their real experience. In the end, the researchers should analyze the data collected from each subject to infer about its experience level. To support the characterization through $R^3$, the authors recommend taking preference to interviews rather than questionnaires. However, one can see that following this strategy may be unfeasible according to the nature of the sample and its size. 

Different from the previous schemes proposed, our research aim at reaching a scheme combining few quantitative and qualitative attributes for reaching a more comprehensive characterization of experience. Besides, we intend to compose a characterization scheme composed of items easily gathered through characterization forms, avoiding the conduction of interviews. Finally, our research intends to support researchers in internally ranking and grouping subjects in empirical studies rather than classifying them through absolute levels of experience.

In this way, an important background from our research includes a previous investigation on the attributes used for characterizing subjects in surveys from the field~\cite{de2015characterizing}. For this purpose, it was conducted a literature review over surveys published in Empirical Software Engineering conferences until 2014. In these sources, they found 38 surveys published. After analysing their content, it was found that most of these studies (32) have individuals as the unit of analysis. From these, the most common information gathered from the subjects is their experience in the research topic. The indicators used for supporting this characterization include the experience level working in the research topic, number of projects, and number of publications. The authors recommend that the subjects' experience should be characterized by combining the overall experience in Software Engineering with the experience in the research topic. Besides, the characterization of subjects is one of the main activities of the conceptual framework proposed in~\cite{de2016surveys}. In this work, it is recognized the lack of standardization on characterizing the subjects' experience. For mitigating this issue, it is recommended the adoption of the more frequent characteristics found in~\cite{de2015characterizing}.

\section{Study Design}
\label{sec:design}

Our research goal aims at characterizing an evidence-based scheme for supporting a comprehensive characterization of the experience of subjects from software engineering studies. By scheme, we define a set of attributes that complementary characterize the experience of each study subject. These attributes may be quantitative and qualitative ones but should be gathered through characterization forms. Besides, the scheme's attributes should be easily adapted for characterizing experience in different skills. The main expected benefit of our research is providing a hands-on resource to support the ranking and grouping of individuals in empirical studies from the field by their experience. Based on this goal and the more common perspectives/attributes observed in previous work (Section ~\ref{sec:relatedwork}), we designed the first version of our characterization scheme, composed of the key attributes presented in Table\ref{tab:scheme}. 

\setlength{\tabcolsep}{10pt} 
\renewcommand{\arraystretch}{1.1} 

\begin{table}[ht!]
	\centering
	\caption{Schema for Characterizing the Experience of Subjects.}
	\label{tab:scheme}
    \begin{tabular}{|c|l|c|}
    
    \hline
\textbf{Perspective of} & \textbf{Attribute} & \textbf{Data Type} \\
\textbf{Experience} &  &  \\
\hline
Time & Years of experience in the topic & Numeric \\   \hline
Projects & Number of Projects in the topic & Numeric \\  \hline
Self-assessment & Self-assessment of the experience & Likert scale \\
& in the topic & (Very Low, Low,\\
& & High, Very High) \\
\hline
\end{tabular}
\end{table}

One may see that \textit{years of experience} and \textit{number of projects} are quantitative attributes commonly used in empirical studies from the field. The self-assessment of experience intends to be a measurable qualitative attribute for gathering complementary and subjective aspects of experience. Therefore, our scheme is based on primitive measures for ranking subjects from the same study sample. In this way, we avoid arbitrarily establishing absolute categories or levels of experience once we believe they should be identified in the context of each study when needed. In the following subsections, we present the design of the empirical study conducted for evaluating the proposed scheme.  
\subsection{Research Questions}

Considering our research goal and the proposed scheme, in this study we intend to answer the following research questions:\\


RQ1. \textit{To which extent the different attributes used for ranking the experience of subjects from Software Engineering studies are correlated?}\\

RQ2. \textit{Which attributes subjects from Software Engineering studies spontaneously adopt for self-describing their experience in software development?} \\

By answering RQ1, we want to observe the extent to which measuring subjects' experience by each attribute results in similar rankings. If two attributes result in highly similar rankings, they are probably redundant. Consequently, we could discard one of them could from the scheme proposed. By answering RQ2, we want to identify relevant perspectives that may be added to evolve the proposed scheme.

\subsection{Population and Sample}
Considering the goal of our research, the target population includes any software developer ranging from novice to very experienced ones. Once we need to gather characterization data, we opportunistically choose a sampling frame composed of participants from four empirical studies in Software Engineering conducted from 2017 until 2020. In all these studies, the subjects were invited by convenience. They are from different organizations and universities from Brazil, among students and practitioners. Two of these studies are controlled experiments on the manual identification of code smells~\cite{de2017influence}~\cite{oliveira2017collaborative}. A third study investigated the social representations of the identification of code smells ~\cite{de2019CHASE}. The fourth work is a controlled study investigating the incidence of confusing code~\cite{de2020atoms}. One can see that the research topics of these investigations address code review tasks.   

\subsection{Instrumentation}
We applied a standard set of characterization items for experience in the four studies, summarized in Table \ref{tab:items}. These characterization questions aim at gathering experience data from the perspective of the scheme's attributes for the following three activities: \textit{software development in general}, \textit{programming in a particular language} and \textit{code reviews}. We intentionally opted by investigating these activities once they address different characterization challenges. Software development is a considerably generic activity. Programming in a particular language is a more specific activity, although probably experienced by most of the subjects. On the other hand, the experience in code reviews may be concentrated in particular subgroups. We included the self-description question about the experience in software development to support answering RQ2.

\setlength{\tabcolsep}{10pt} 
\renewcommand{\arraystretch}{1.1} 

\begin{table}[ht!]
	\centering
	\caption{Characterization items for experience.}
	\label{tab:items}
    \begin{tabular}{|c|l|c|}
    
    \hline
\textbf{Perspective} & \textbf{Questionnaire Item} & \textbf{Data Type} \\
\hline
Self & Summary of experience with software & Open \\   
description& development&\\ \hline
 & Self-perception of experience with software & Likert scale\\ 
 & development &\\
Self- & Self-perception of experience with (C/Java) & Likert scale \\
assessment&Self-perception of experience with code & Likert scale \\ 
& reviews & \\ \hline
&Years developing software  & Numeric\\                   
Time & Years programming in (C/Java)  & Numeric\\                   
&Years reviewing code & Numeric\\ \hline  

&Number of projects developing software& Numeric\\      
Projects &Number of projects programming in & Numeric\\  
& (C/Java) &\\
&Number of software projects reviewing code & Numeric\\\hline

\end{tabular}
\end{table}

Besides the presented set characterization items, other ones were eventually added to address the scope of each study. To avoid bias, we designed each characterization form to gather data about each attribute in separated sections by following the same order presented in Table \ref{tab:items}.

\subsection{Data Analysis}

After filtering the answers given from 79 subjects, we calculated the coefficient of correlation among the different attributes for each activity investigated (RQ1). Considering the heteroscedasticity and the sample size, we opted by applying Kendall's non-parametric correlation test~\cite{kendall} for each of the following pairs of distribution: \textit{self-assessment x years of experience; self-assessment x number of projects; years of experience x number of projects}. For answering RQ2, we performed a single level of open coding over the self-description answers for identifying the attributes used by developers to describe their experience. Then, we ranked these attributes by frequency.

\section{Results}
\label{sec:results}

Tables ~\ref{tab:statisticsSD},~\ref{tab:statisticsPG}, and ~\ref{tab:statisticsCR} summarize the descriptive statistics of the experience attributes measured from the 79 subjects for each activity investigated.

\setlength{\tabcolsep}{10pt} 
\renewcommand{\arraystretch}{1.1} 

\begin{table}[ht!]
	\centering
	\caption{Descriptive statistics of the developers' experience with software development.}
	\label{tab:statisticsSD}
    \begin{tabular}{|l|c c c|}
    \hline
\textbf{Item} & \textbf{Self} & \textbf{Years} & \textbf{Projects} \\
 & \textbf{Assessment} &  &  \\
\hline
Minimum     & 0.00    & 0.00      & 0.00   \\
1st Quartile& 2.00    & 3.50      & 4.00  \\
Median      & 2.00    & 5.00      & 7.00 \\ 
Mean        & 1.91 & 7.25   & 11.25 \\
3rd Quartile& 2.00   & 8.50      & 13.00  \\ 
Maximum     & 3.00   & 40.00     & 50.00   \\ \hline
\end{tabular}
\end{table}

\begin{table}[ht!]
	\centering
	\caption{Descriptive statistics of the developers' experience with programming in a particular language.}
	\label{tab:statisticsPG}
    \begin{tabular}{|l|c c c|}
    \hline
\textbf{Item} & \textbf{Self} & \textbf{Years} & \textbf{Projects} \\
 & \textbf{Assessment} &  &  \\
\hline
Minimum     & 0.00    & 0.00      & 0.00   \\
1st Quartile& 1.00    & 1.00      & 1.00  \\
Median      & 2.00    & 3.00      & 3.00 \\ 
Mean        & 1.53 & 3.93   & 4.90 \\
3rd Quartile& 2.00    & 5.50      & 5.00  \\ 
Maximum     & 3.00    & 20.00     & 30.00   \\ \hline
\end{tabular}
\end{table}

\begin{table}[ht!]
	\centering
	\caption{Descriptive statistics of the developers' experience with code reviews.}
	\label{tab:statisticsCR}
    \begin{tabular}{|l|c c c|}
    \hline
\textbf{Item} & \textbf{Self} & \textbf{Years} & \textbf{Projects} \\
 & \textbf{Assessment} &  &  \\
\hline
Minimum     & 0.00    & 0.00      & 0.00   \\
1st Quartile& 0.50    & 0.00      & 0.00  \\
Median      & 1.00    & 2.00      & 2.00 \\ 
Mean        & 1.35 & 3.03   & 4.54 \\
3rd Quartile& 2.00    & 4.50      & 5.00  \\ 
Maximum     & 3.00    & 20.00     & 40.00   \\ \hline
\end{tabular}
\end{table}

The data about subjects' overall experience in software development in terms of years and projects indicates that these subjects range from novice developers to very experienced ones. Besides, the distribution of self-assessment reveals a trend on most of these developers assessing themselves as having a high level of experience (median=2, mean=1.91).

The data about subjects' experience in a particular programming language reveals the distribution is considerably diverse in terms of time and number of projects. Besides, one can see the concentration of novices in the first quartile (0 to 1 years, 0 to 1 project). However, the distribution of self-assessment reveals a trend on these developers assessing themselves as having a high level of experience in this activity (median=2, mean=1.53). Finally, the distributions of experience in code reviews indicate an even higher concentration of novices in the first quartile (0 years, 0 projects), reflected in the self-assessment.

The distributions of experience analysed indicate that the study sample is considerably diverse for the different activities investigated. Besides, this diversity can be observed among the different attributes from the proposed scheme. Therefore, we understand the collected dataset is favourable for supporting answering RQ1, which is discussed in the following subsection. RQ2 is answered in Section \ref{subsec:selfdescription}

\subsection{RQ1: Comparing the Characterization Attributes}

By answering RQ1, we intend to observe to which extent the different attributes used for ranking developers are correlated. Tables \ref{tab:correlationSD}, \ref{tab:correlationPL}, and \ref{tab:correlationCR} present the Kendall's correlation coefficients calculated for each pair of distributions evaluated. In total, we performed nine correlation tests involving the 711 answers given by the 79 subjects. In all the correlation tests performed, the p-values obtained were lesser than 0.0001.  

\setlength{\tabcolsep}{10pt} 
\renewcommand{\arraystretch}{1.1} 

\begin{table}[ht!]
\centering
\caption{Kendall's correlation coefficients calculated for experience with software development.}
\label{tab:correlationSD}
\begin{tabular}{|c| c c c|}
    \hline
    \textbf{Attribute} & \textbf{Self} & \textbf{Years} & \textbf{Projects} \\
    \hline
    \textbf{Self}             & -   &       &  \\
    \textbf{Years}           & 0.5620    & -      &  \\
    \textbf{Projects}      & 0.4972    & 0.5601      & - \\  \hline
\end{tabular}
\end{table}

\begin{table}[ht!]
\centering
\caption{Kendall's correlation coefficients calculated for experience with a particular programming language.}
\label{tab:correlationPL}
\begin{tabular}{|c|c c c|}
    \hline
    \textbf{Attribute} & \textbf{Self} & \textbf{Years} & \textbf{Projects} \\
    \hline
    \textbf{Self}             & -   &       &  \\
    \textbf{Years}          & 0.6502    & -      &  \\
    \textbf{Projects}      & 0.6929    & 0.6616      & - \\  \hline
\end{tabular}
\end{table}

\begin{table}[ht!]
\centering
\caption{Kendall's correlation coefficients calculated for experience with code reviews.}
\label{tab:correlationCR}
\begin{tabular}{|c|c c c|}
    \hline
    \textbf{Attribute} & \textbf{Self} & \textbf{Years} & \textbf{Projects} \\
    \hline
    \textbf{Self}            & -   &       &  \\
    \textbf{Years}           & 0.6705    & -      &  \\
    \textbf{Projects}      & 0.6486    & 0.8362      & - \\  \hline
\end{tabular}
\end{table}

Considering the nature of the study dataset (human aspects), we used the categorisation proposed by Dancey and Raidy~\cite{dancey_reidy} for interpreting the coefficients of correlation obtained. This categorization is reproduced in Table \ref{tab:correlation}. Based on this, one can see that almost all Kendall's correlation coefficients calculated indicate moderately positive correlations among the attributes compared. The only exception is the strong correlation obtained for \textit{years x projects} in \textit{code reviews}. Based on these results, we may depict the following main finding:
\\\\
\fbox{
  \parbox{\textwidth}{\textbf{Finding 1.} \textit{The experience of developers in terms of time, projects, and \\ self-assessment are positively but moderately correlated.}}\\\\
}

\begin{table}[ht!]
\centering
\caption{Interpretation proposed for Kendall's positive correlations. Adapted from~\cite{dancey_reidy}.}
\label{tab:correlation}
\begin{tabular}{|c c|}
    \hline
    \textbf{Kendall's tau} & \textbf{Correlation} \\ \hline
    1         & perfect \\
    0.7000-0.9999  & strong \\
    0.4000-0.6999 & moderate \\
    0.1000-0.3999 & weak \\
    0-0.0999    & none \\ \hline
\end{tabular}
\end{table}
 
Besides, based on Finding 1 and the nature of Kendall's test, we may depict the following main finding:\\

\fbox{%
  \parbox{\textwidth}{
    \textbf{Finding 2.} \textit{Ranking subjects' experience exclusively by years of experience, \\
    projects, or self-assessment leads to different results.}}\\
}

\subsection{RQ2: Self-Description of Experience} \label{subsec:selfdescription}

For answering RQ2, we asked the subjects to summarize their experience with software development. The answers resulted in 1,646 words. We performed a single round of open coding in these words to identify the set of attributes used by the developers for self-describing their experience. From the 79 respondents, 59 developers reported at least one characteristic of background. Table \ref{tab1} shows the nine attributes coded, with their incidence among the 59 answers coded. 

\setlength{\tabcolsep}{10pt} 
\renewcommand{\arraystretch}{1.1} 

\begin{table}
\centering
\caption{Frequency (f) of the attributes used by developers to self-describing their experience in software development.}\label{tab1}
\begin{tabular}{|l | r | r|}
\hline
\textbf{Attribute} & \textit{f} & \textbf{\%} \\
\hline
Time (months/years) & 30 & 50.85\% \\
Enumeration of Programming Languages & 27 & 45.76\% \\
Enumeration of Development Activities & 16 & 27.12\% \\
Enumeration of Software Technologies & 13 & 22.03\% \\
Particular Research Projects & 9 & 15.25\% \\
Enumeration of Project Domains & 7 & 11.86\% \\
Particular Working Companies & 6 & 10.17\% \\
Number of Projects & 3 & 5.08\% \\
Experience with Entrepreneurship & 2 & 3.39\% \\
\hline
\end{tabular}
\end{table}

One can see that \textit{time} was the single attribute mentioned by more than half of answers coded (30). However, the experience with \textit{programming languages} was also largely mentioned (27). Curiously, the \textit{number of projects}, another quantitative attribute from our scheme, was barely mentioned (3). The report of particular research projects (9) may be explained due several subjects investigated are also post-graduation students or experienced researchers. Most of them had worked with development activities at one or more research projects. 

Different attributes revealed the concern of developers on enumerating the technical diversity of their software development experience. By \textit{development activities} (16), we mean the cases in which developers listed activities they are used to perform, including different software engineering disciplines, such as requirements, design, implementation, and verification. In these cases, most of the developers listed two or more activities. Similarly, several developers reported their experience in terms \textit{programming languages} (27) and other\textit{software technologies} (13) adopted. These technologies include particular frameworks, APIs, and operating systems. Therefore, the aforementioned results lead us to depict the following main findings:\\

\fbox{%
  \parbox{\textwidth}{
    \textbf{Finding 3.} \textit{Software developers tend to quantitatively characterize their\\
    experience with software development in terms of time rather than in number\\
    of projects.}}\\
}

\fbox{%
  \parbox{\textwidth}{
    \textbf{Finding 4.} \textit{Software developers tend to enumerate their experience with\\
    software development through the diversity of their technical backgrounds.}}\\
}

\section{Discussion}
\label{sec:discussion}

The findings of our study indicates that both quantitative and qualitative sources of knowledge are relevant to reach a more comprehensive and accurate characterization of developers' experience. One could argue that once more attributes are adopted, the more comprehensive would be the characterization. However, fulfilling long characterization forms tends to be unfeasible when running studies in the field, especially when involving professional developers out of the researchers' network. Long questionnaires may discourage individual participation or even discourage potential supporters and sponsors, such as managers and local research and development representatives. Besides, applying a large number of questions- especially open ones- would demand considerable analysis effort from researchers, which could be unfeasible for large samples. Therefore, we should reach an optimized scheme for gathering relevant characterization data as much as possible.

In the first version of our schema, we combined two quantitative attributes (years and number of projects) with a single qualitative attribute (self-assessment), gathered through a standard Likert scale. The study findings indicate that these attributes effectively bring complementary perspectives for characterizing experience, especially those experience in more generic activities, such as software development. However, it is not clear to which extent they are sufficient. For instance, are there other attributes that worth composing the schema? On the other extreme, are self-assessments sufficient to synthesize all the subject experience?

Despite the exploratory nature of this first study, the analysis of the self-description answers (open questions) gives us some important hints for answering these questions. First, no other quantitative attribute than time was considerably reported. Second, the perception of technical diversity also emerged as a potentially relevant attribute. In this way, one possible direction for evolving the original scheme would be replacing the number of projects with self-assessment questions addressing the diversity of projects in terms of their domains, relevant technologies, and relevant activities. 

However, the study findings indicate that we should perform a deeper investigation into what researchers and developers have in mind about the research topic before evolving the schema. In this sense, we are planning a new study using the theory of social representations from Social Psychology to support our research. This theory aims to establish an order that enables how groups of individuals guide themselves in their material and social world. Social representation is the collective elaboration of a social object by a particular community for behaving and communicating~\cite{moscovici_1988}. In this definition, a \textit{social object} corresponds to an object socialized by two or more individuals from a community, such as the experience of software developers. Recent studies have shown the usefulness of the social representation theory for supporting both research and practice in software engineering ~\cite{de2019CHASE}~\cite{de2019ESEM}~\cite{de2021CHASE}. 

\section{Threats to Validity}
\label{sec:threats}

Different characterization forms were designed to supporting each empirical study used in this investigation. This issue represents an important threat to the construct validity, once additional questions would influence the developers' answers. To mitigate this threat, we distributed the scheme items as presented in Table \ref{tab:items} at the beginning of each characterization form. Besides, we used proper settings to assure that developers would answer the questions addressing only a single perspective at a time.

Regarding the external validity, one can see that the four studies involved in this investigation address code reviews. Therefore, one may argue that the findings may do not apply to other research topics. However, it is important to note that we also characterized developers through a generic skill, i.e., software development. Besides, Section \ref{sec:results} shows that developers with different levels of experience and diverse skills participated in the study. Therefore, not only specialists in code reviews composed the study sample.

\section{Conclusion and Future Work}
\label{sec:conclusion}
Software Engineering research has several challenges addressing the support to empirical studies. One important challenge addresses the proper characterization of the studies' subjects, especially about their experience. The characterization of this experience may play a key role in the study design and in the interpretation of its results.

The scheme proposed in this paper for supporting the characterization of study subjects intends to be comprehensive but also simplified and easy to use for both researchers and subjects. The results of the empirical study reported in this paper suggest that our schema is feasible, but its comprehensiveness may be improved. However, it is still not clear the concrete changes that we should make. In this sense, we are designing an empirical study to investigate what researchers and developers have in mind about experience. We believe that this study would help us reach a more comprehensive characterization scheme still preserving its simplicity.

\section{Acknowledgements}
\label{sec:acknowledgements}
This work is funded by CNPq 152179/2020-8.

%
%
%
\bibliographystyle{splncs04}
\bibliography{CIBSE_characterizing}

\begin{thebibliography}{10}
\providecommand{\url}[1]{\texttt{#1}}
\providecommand{\urlprefix}{URL }
\providecommand{\doi}[1]{https://doi.org/#1}

\bibitem{dancey_reidy}
Dancey, C.P., Reidy, J.: Statistics without maths for psychology. Pearson
  education (2007)

\bibitem{dybaa2012works}
Dyb{\aa}, T., Sj{\o}berg, D.I., Cruzes, D.S.: What works for whom, where, when,
  and why? on the role of context in empirical software engineering. In:
  Proceedings of the ACM-IEEE international symposium on Empirical software
  engineering and measurement. pp. 19--28 (2012)

\bibitem{falessi2018empirical}
Falessi, D., Juristo, N., Wohlin, C., Turhan, B., M{\"u}nch, J., Jedlitschka,
  A., Oivo, M.: Empirical software engineering experts on the use of students
  and professionals in experiments. Empirical Software Engineering
  \textbf{23}(1),  452--489 (2018)

\bibitem{host2000using}
H{\"o}st, M., Regnell, B., Wohlin, C.: Using students as subjects—a
  comparative study of students and professionals in lead-time impact
  assessment. Empirical Software Engineering  \textbf{5}(3),  201--214 (2000)

\bibitem{host2005experimental}
H{\"o}st, M., Wohlin, C., Thelin, T.: Experimental context classification:
  incentives and experience of subjects. In: Proceedings of the 27th
  international conference on Software engineering. pp. 470--478 (2005)

\bibitem{kendall}
Kendall, M.G.: A new measure of rank correlation. Biometrika  \textbf{30}(1/2),
   81--93 (1938)

\bibitem{kontio2008focus}
Kontio, J., Bragge, J., Lehtola, L.: The focus group method as an empirical
  tool in software engineering. In: Guide to advanced empirical software
  engineering, pp. 93--116. Springer (2008)

\bibitem{lenberg2015behavioral}
Lenberg, P., Feldt, R., Wallgren, L.G.: Behavioral software engineering: A
  definition and systematic literature review. Journal of Systems and software
  \textbf{107},  15--37 (2015)

\bibitem{linaker2015guidelines}
Lin{\aa}ker, J., Sulaman, S.M., de~Mello, R.M., H{\"o}st, M.: Guidelines for
  conducting surveys in software engineering. Lund Univeristy  (2015)

\bibitem{de2021CHASE}
de~Mello, R., da~Costa, J.A., de~Oliveira, B., Ribeiro, M., Fonseca, B., Gheyi,
  R., Garcia, A., Tiengo, W.: Decoding confusing code: Social representations
  among developers. In: 2021 IEEE/ACM 13th International Workshop on
  Cooperative and Human Aspects of Software Engineering (CHASE). pp. 11--20.
  IEEE (2021)

\bibitem{de2019CHASE}
de~Mello, R., Gonçalves~Uchoa, A., Felicio~Oliveira, R., Tenório Martins~de
  Oliveira, D., Fonseca, B., Fabricio~Garcia, A., de~Barcellos~de Mello, F.:
  Investigating the social representations of code smell identification: A
  preliminary study. In: 2019 IEEE/ACM 12th International Workshop on
  Cooperative and Human Aspects of Software Engineering (CHASE). pp. 53--60
  (2019). \doi{10.1109/CHASE.2019.00022}

\bibitem{de2019ESEM}
de~Mello, R., Uchôa, A., Oliveira, R., Oizumi, W., Souza, J., Mendes, K.,
  Oliveira, D., Fonseca, B., Garcia, A.: Do research and practice of code smell
  identification walk together? a social representations analysis. In: 2019
  ACM/IEEE International Symposium on Empirical Software Engineering and
  Measurement (ESEM). pp.~1--6 (2019). \doi{10.1109/ESEM.2019.8870141}

\bibitem{de2017influence}
de~Mello, R.M., Oliveira, R., Garcia, A.: On the influence of human factors for
  identifying code smells: A multi-trial empirical study. In: 2017 ACM/IEEE
  International Symposium on Empirical Software Engineering and Measurement
  (ESEM). pp. 68--77. IEEE (2017)

\bibitem{de2015characterizing}
de~Mello, R.M., Travassos, G.H.: Characterizing sampling frames in software
  engineering surveys. In: Proceedings of the XVIII IberoAmerican Conference on
  Software Engineering (CIbSE). p.~81 (2015)

\bibitem{de2016surveys}
de~Mello, R.M., Travassos, G.H.: Surveys in software engineering: Identifying
  representative samples. In: Proceedings of the 10th ACM/IEEE International
  Symposium on Empirical Software Engineering and Measurement. pp.~1--6 (2016)

\bibitem{moscovici_1988}
Moscovici, S.: Notes towards a description of social representations. European
  Journal of Social Psychology  \textbf{18},  211 -- 250 (07 1988).
  \doi{10.1002/ejsp.2420180303}

\bibitem{de2020atoms}
de~Oliveira, B., Ribeiro, M., da~Costa, J.A.S., Gheyi, R., Amaral, G.,
  de~Mello, R., Oliveira, A., Garcia, A., Bonif\'{a}cio, R., Fonseca, B.: Atoms
  of confusion: The eyes do not lie. p. 243–252. SBES '20, Association for
  Computing Machinery, New York, NY, USA (2020). \doi{10.1145/3422392.3422437},
  \url{https://doi.org/10.1145/3422392.3422437}

\bibitem{oliveira2017collaborative}
Oliveira, R., Sousa, L., de~Mello, R., Valentim, N., Lopes, A., Conte, T.,
  Garcia, A., Oliveira, E., Lucena, C.: Collaborative identification of code
  smells: A multi-case study. In: 2017 IEEE/ACM 39th International Conference
  on Software Engineering: Software Engineering in Practice Track (ICSE-SEIP).
  pp. 33--42. IEEE (2017)

\bibitem{runeson2009guidelines}
Runeson, P., H{\"o}st, M.: Guidelines for conducting and reporting case study
  research in software engineering. Empirical software engineering
  \textbf{14}(2),  131--164 (2009)

\bibitem{salman2015students}
Salman, I., Misirli, A.T., Juristo, N.: Are students representatives of
  professionals in software engineering experiments? In: 2015 IEEE/ACM 37th
  IEEE International Conference on Software Engineering. vol.~1, pp. 666--676.
  IEEE (2015)

\bibitem{dos2009action}
dos Santos, P.S.M., Travassos, G.H.: Action research use in software
  engineering: An initial survey. In: 2009 3rd International Symposium on
  Empirical Software Engineering and Measurement. pp. 414--417. IEEE (2009)

\end{thebibliography}

\end{document}